\documentclass{ws-ijmpe}

\begin{document}

\markboth{C.W. Shen, Y. Abe, D. Boilley, G. Kosenko, E.G. Zhao}
{Isospin dependence of reactions $^{48}$Ca+$^{243-251}$Bk}

\catchline{}{}{}{}{}
%%%%%%%%%%%%%%%%%%%%%%%%%%%%%%%%%%%%%%%%%%%%%%%%%%%%%%%%%%%%%%%%%%%%

\title{Isospin dependence of reactions $^{48}$Ca+$^{243-251}$Bk}

\author{CAIWAN SHEN}

\address{School of Science, Huzhou Teachers College, \\
    Huzhou 313000, Zhejiang Province, P.R. China \\
    cwshen@hutc.zj.cn }

\author{YASUHISA ABE}
\address{Research Center for Nuclear Physics, Osaka University, 10-1 Mihogaoka, \\
         Ibaraki, 567-0047/Osaka, Japan \\
         abey@rcnp.osaka-u.ac.jp  }

\author{DAVID BOILLEY}
\address{GANIL, CEA/DSM-CNRS/IN2P3, BP 55027, F-14076 Caen cedex 5, France \\
    and Univ. of Caen, B.P. 5186, F-14032 Caen Cedex, France \\ 
    boilley@ganil.fr }

\author{GRIGORY KOSENKO}
\address{Department of Physics, Omsk University, Omsk, 644077, Russia \\ 
    kosenko@phys.omsu.omskreg.ru}

\author{ENGUANG ZHAO}
\address{Institute of Theoretical Physics, Chinese Academy of Science, \\
    Beijing 100080, P.R. China \\
    egzhao@itp.ac.cn }

\maketitle

\begin{history}
\received{(received date)}
\revised{(revised date)}
%\accepted{(Day Month Year)}
%\comby{(xxxxxxxxxx)}
\end{history}

\begin {abstract}
The fusion process of $^{48}$Ca induced reactions is studied with the two-step 
model. In this model, the fusion process is devided into two stages: first, 
the sticking stage where projectile and target come to the touching point over 
the Coulomb barrier from infinite distance, and second, the formation stage 
where the di-nucleus formed with projectile and target evolve to form the 
spherical compound nucleus from the touching point. By the use of the 
statistical evaporation model, the residue cross sections for different neutron
evaporation channels are analyzed. From the results, optimum reactions are given
to synthesize $Z$ = 117 element with $^{48}$Ca induced reactions. 
\end{abstract}

\section{Introduction}

Since the time when the nuclear shell model was constructed, the question -- 
where the next double magic nucleus beyond $^{208}$Pb is -- has always attracted 
physicists' attention.  It is expected to have a long life time due to 
the additional stability by the closed shell.  Based on the shell model with 
a single particle potential, the next double magic nucleus is predicted to 
be of 114 protons and 184 neutrons. Other theories predict that the next 
magic proton number may also be 120, 126.  Thus, the location of the next 
double magic nucleus in the nuclear chart is model- or even parameter-dependent. 
In order to prove an existence of such elements, we have to 
synthesize them, since  there is no sure evidence of such elements in the 
nature until now.  How to reach them is still very difficult experimentally.  

With the development of experimental facilities, the minimum cross section
that could be detected is about picobarn or even lower and accordingly many
superheavy elements (SHE) with proton number $Z\geq100$ were discovered.
Among the discovered SHE, the decay chains are not always connected with known nuclei. 
Therefore, where the magic nucleus is located and how to reach it is still an
open question. SHE events have been detected for $Z=114-116,118$\cite{oganessian}, 
except for $Z=117$. For those SHE where $Z\geq114$, the so-called hot fusion reaction
with $^{48}$Ca as a projectile is adopted.

The neighbors of $Z$ = 117 are synthesized with projectile $^{48}$Ca, then it is natural
to think that $Z$ = 117 can also be synthesized with the same projectile. If it is 
true, the optimum target and incident energy should be suggested.
However, because of the complexity of heavy ion fusion processes, 
no commonly accepted reaction theory is available.
Nevertheless,  several theoretical attempts have been made to explain or predict the
experiments, for example, the two-step model which will be used in this 
paper\cite{cwshen}, DNS model\cite{dns}, QMD-based model\cite{qmd}, and the other 
two models in Ref.\cite{swiatecki2005,jdbao}. A common problem faced by these theories is the
hindrance of fusion, which is simply characterized by so-called extra-push energy in addition to 
the Coulomb barrier energy, i.e., the additional energies needed in order to form a compound nucleus 
in the heavy ion collisions.  For theoretical prediction of fusion and thus 
residue cross sections, an understanding of the mechanism of the hindrance is 
indispensable.  

Recently, the present authors et al. have proposed a mechanism  
where the main origin of the hindrance is due to a relative 
position between the di-nucleus configuration formed by the sticking of 
incident ions and the conditional saddle point or more generally the ridge 
line in the liquid drop energy (LDM) energy surface of the composite system\cite{abe-eurojp}. 
The model explains the extra-push energy and furthermore gives an energy 
dependent fusion probability\cite{abe2000}. The two-step 
model is based on the mechanism, including effects due to collision processes 
over the Coulomb barrier, which will be briefly recapitulated below. 

The purpose of the present paper is to provide a reliable prediction for synthesis 
of the new superheavy element with $Z$ = 117, that is, to suggest most promising target 
isotopes of berkelium and to predict excitation functions for $x$n residue cross sections 
which give an optimum $E_{\rm c.m.}$ and a peak value of residue cross sections.  
For that, we employ the two-step model for fusion process and a part of 
{\small HIVAP} code\cite{reisdorf} for the decay  process, where a calibration of 
only one free parameter is made by the use of the neighboring
reactions with the target $^{244}$Pu and $^{248}$Cm. The paper is arranged as follows: 
Sec. 2 describes the two-step model used in the description of fusion process; Sec. 3 
shows the results of calculations and discussions; Sec. 4 gives a summary.

\section{Two-step model}

Based on the theory of compound nucleus reactions, the residue cross sections
are given as follows:
\begin{equation}
\sigma_{\rm res}=\pi\lambda{\hspace{-2.3mm}{^-}}^{2}\sum_{J}(2J+1)P_{\rm fusion}
^{J}(E_{\rm c.m.})\cdot P_{\rm surv}^{J}(E^{\ast}), \label{eq-1}
\end{equation}
where $\lambda{\hspace{-2.3mm}{^-}}=\lambda/(2\pi)=\hbar/\sqrt{2\mu E_{\rm c.m.}}$, $J$ is
the total angular momentum quantum number and $\mu$ 
the reduced mass. $P_{\rm fusion}$ and
$P_{\rm surv}$ denote the fusion and the survival probabilities,
respectively. The latter one is given by the statistical theory,
although there are ambiguities in the parameters in the properties of heavy and 
superheavy nuclei, which give rise to uncertainties in calculating the residue 
cross sections. In addition, the first one, $P_{\rm{fusion}}$, the fusion 
probability of massive systems, is essentially unknown.  The reason is the 
so-called fusion hindrance in heavy ion collisions, which is experimentally 
well known, but is not understood in its mechanism. In lighter systems, the fusion 
probability is well determined by the Coulomb repulsion and nuclear attraction 
between projectile and target, while in massive systems fusion does not occur 
in the same way. One must give an additional incident energy (extra-push energy) 
to explain the data. There are two interpretations trying to explain the phenomenon: 
one is due to the dissipation of the initial kinetic energy during two-body 
collisions passing over the barrier\cite{frobrich}, and the other one is due to 
the dissipation of the energy of collective motions which would lead an amalgamated 
system to the spherical compound nucleus\cite{swiatecki1981}. Because the two 
mechanisms are considered in different stages of the fusion process, the two 
mechanisms can be combined into a single model, the two step model\cite{cwshen,abe2002}, 
in which energy dissipation takes place in both stage: the overcoming of Coulomb 
potential before touching point, called approaching phase, and the evolution 
of the amalgamated system after touching point\cite{swiatecki1981}, called formation 
phase. Because they are the successive processes, they should be connected.  
The method of statistical connection\cite{boilley2003} has been proposed in 
the two step model, which will be explained later in the application.

In the description of approaching phase, we have two options: one is that the
phase may be described as collision process under frictional forces, as done
in Ref.\cite{cwshen}; the other choice is to adopt an empirical formula
to reproduce the experimental capture cross sections and then extend it to unknown
region\cite{wilczynska}. In the formation phase, we describe dynamical
evolution of the amalgamated mononuclear system toward the spherical shape
under frictional forces acting on collective motions of excited nuclei. For
each angular momentum $J$, the sticking probability $P_{\rm stick}^{J}$, 
i.e. the probability from infinite distance to the contact point, and
formation probability $P_{\rm form}^{J}$, i.e., the probability from the
contact point to the spherical compound nucleus, can be worked out and then
fusion probability and fusion cross section get the form,
\begin{equation}
P_{\rm fusion}^{J}(E_{\rm c.m.})=P_{\rm{stick}}^{J}(E_{\rm{c.m.}
})\cdot P_{\rm{form}}^{J}(E_{\rm{c.m.}}), \label{eq-2}
\end{equation}
and
\begin{equation}
\sigma_{\rm{fusion}}(E_{\rm{c.m.}})=\pi\lambda{\hspace{-2.3mm}{^-}}^{2}
{\displaystyle\sum\limits_{J}}
(2J+1)P_{\rm{fusion}}^{J}(E_{\rm{c.m.}}), \label{eq-22}
\end{equation}
respectively.

\subsection{Approaching phase}

\begin{figure} [h]
\centerline{\psfig{file=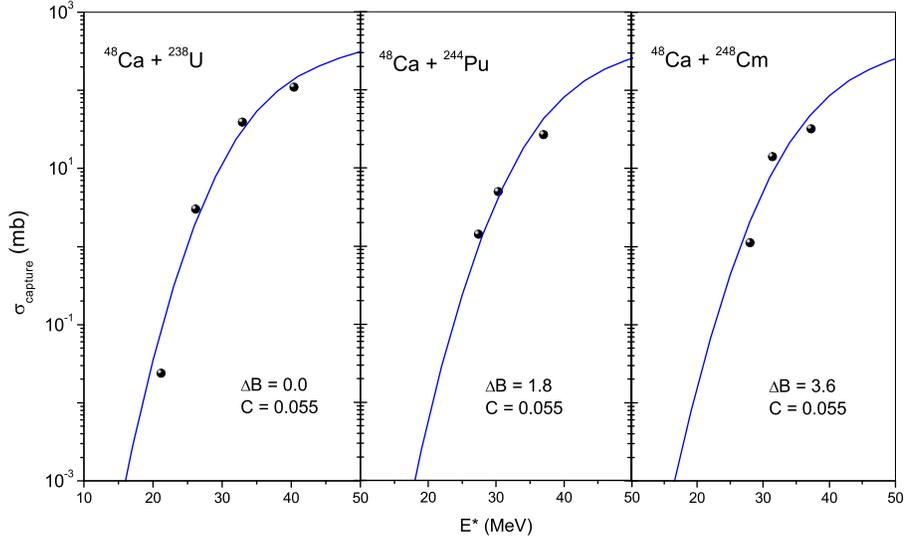,width=12cm}}
\vspace*{8pt}
\caption{Fit of the experimental capture cross section to get appropriate $C$ and $\Delta B$.}
\label{fig1}
\end{figure}

\begin{figure} [h]
\centerline{\psfig{file=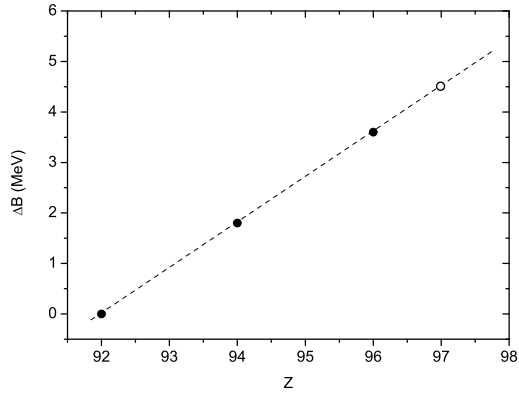,width=6.8cm}}
\vspace*{8pt}
\caption{Extrapolation of the shift of Coulomb barrier for Ca + Bk. The solid
circles correspond to experimental data, while the open one is an extrapolation for Ca+Bk.}
\label{fig2}
\end{figure}

In this phase, following Eq.(4) in Ref.\cite{swiatecki2005} and under the 
assumption of $B_{0}/(\sqrt{2}H)\gg 1$, the capture probability for each 
partial wave is given as
\begin{equation}
P_{\rm{stick}}^{J}(E_{\rm{c.m.}})=\frac{1}{2}\left\{  1+\operatorname{erf}\left[ \frac
{1}{\sqrt{2}H}\left(  E_{cm}-B_{0}-\frac{\hbar^{2}J(J+1)}{2\mu R_{B}^{2}
}\right)  \right]  \right\}  , \label{eq-3}
\end{equation}
where $B_{0}$ is the barrier height of the Coulomb potential, $H$ the width of
the Gaussian distribution of the barrier height, $R_{B}$ the distance between
two centers of projectile and target at the Coulomb barrier.
In the Ref.\cite{swiatecki2005} 45 reactions are used to obtain the values of 
the parameters $H$ and $B_{0}$. However they might not be adequate for very heavy
systems, such as the systems studied in this paper. A reasonable method is that
we use the same form of formulas but with different $C$ (a factor in the 
empirical formula to calculate $H$, see Ref.\cite{swiatecki2005}) and $B_{0}$ to fit the
capture cross sections of systems which are very close to $^{48}$Ca + Bk
reactions. The three $^{48}$Ca-induced systems to be fitted are $^{48}$Ca +
$^{238}$U, $^{48}$Ca + $^{244}$Pu and $^{48}$Ca + $^{248}$Cm \cite{itkis}. Here
barrier height $B_{0}$ in Eq.(\ref{eq-3}) is referred to $B$ of Ref.\cite{swiatecki2005}
and now $B_0$ is,
\begin{equation}
B_{0}=B+\Delta B. \label{eq-4}
\end{equation}
The fitted results are shown in Fig. \ref{fig1}. With a constant value of $C$ 
and a linear increase of the barrier shift $\Delta B$ with proton number $Z$, 
the capture cross sections are very well reproduced. Because 
berkelium has only one more proton in addition to curium, the $\Delta B$ is extrapolated
to be 4.5 MeV, as is seen in Fig. \ref{fig2}. With Eq.(\ref{eq-3}), Eq.(\ref{eq-4}) 
and the re-fitted data $C$ and $\Delta B$, sticking probability $P_{\rm{stick}}^{J}$ 
of $^{48}$Ca + Bk is calculated with confidence. 

%**********************figure 1 *************************

%**********************figure 2 *************************

According to the surface friction model, the relative kinetic energy is completely 
damped at the contact point, and reaches the thermal equilibrium with the heat 
bath. The radial momentum is, thus, Gaussian distributed. This is the initial 
condition for successive process, i.e., in the next formation phase. Thus, the 
method is called statistical connection.

\subsection{Formation phase}

In this process the amalgamated mononuclear system evolves from the contact point
into compound ground state. 
%************By Abe san*************************
In order to describe shapes of the amalgamated system, three parameters are necessary at least.  
In the Two-Center parametrization, they are the distance between two centers $z$, 
the mass asymmetry $\alpha$ , and the neck correction factor $\varepsilon$. 
The first one is defined as a dimensionless parameter as follows,
$$
z=R/R_{0},
$$
where $R$ denotes the distance between the two centers of the harmonic potentials, 
and $R_0$ the radius of the spherical compound nucleus. The second one is 
defined as usual,
$$
\alpha=\frac{A_{1}-A_{2}}{A_{1}+A_{2}},
$$
where $A_1$ and $A_2$ are mass numbers of the constituent nuclei.  The neck correction 
factor is defined by the ratio of the smoothed height at the connection point 
of the two harmonic potentials and that of spike potential. In the description, 
nuclear shapes are defined by equi-potential surface with a constant volume.
For example, $\varepsilon$ = 1.0 means no correction, i.e., di-nucleus shape, while 
$\varepsilon$ = 0.0 means no spike, i.e., flatly connected potential, 
which describes highly deformed mono-nucleus.  
Thus, the $\varepsilon$ describes shape evolution of the compound system from 
di-nucleus to mono-nucleus.  Since we know that the inertia mass for the $\varepsilon$ 
degree of freedom is small, its momentum is expected to be quickly equilibrated, 
compared with the other two degrees of freedom.  And furthermore, LDM potential is rather 
steep with respect to the $\varepsilon$, and then $\varepsilon$ very quickly reaches 
the end, at $\varepsilon$ = 0.0, starting with $\varepsilon$ = 1.0.  
This is natural, considering the strong surface tension of nuclear matter and a sensitive 
change of the surface area due to variation of the $\varepsilon$.  Actually, due to actions 
of the random force associated to the friction, the $\varepsilon$ reaches to 
the equilibrium quickly, far quicker than the time scale of radial fusion 
motion\cite{boilley}.  Thus, we take the $\varepsilon$ = 0.1 (an average value 
in the equilibrium) during the fusion process\cite{abe2008}. The initial parameters 
for $z$ and $\alpha$ are
$$
z_{0}=(A_{p}^{1/3}+A_{t}^{1/3})/(A_{p}+A_{t})^{1/3},
$$
and
$$
\alpha_{0}=(A_{t}-A_{p})/(A_{t}+A_{p}),
$$
respectively.

The evolution of the pear-shaped mono-nucleus after contact point are
described by the multi-dimensional Langevin equations\cite{wada},
\begin{align}
\frac{dq_{i}}{dt}  &  =(m^{-1})_{ij}p_{j},\nonumber\\
\frac{dp_{i}}{dt}  &  =-\frac{\partial U^{J}}{\partial q_{i}}-\frac{1}{2}
\frac{\partial}{\partial q_{i}}(m^{-1})_{jk}p_{j}p_{k}-\gamma_{ij}
(m^{-1})_{jk}p_{k}+g_{ij}R_{j}(t),\label{eq-5}\\
g_{ik}g_{jk}  &  =\gamma_{ij}T^{J},\nonumber
\end{align}
where summation is implicitly assumed over repeated suffixes. In the above
equations, $i$, $j$ takes 1 or 2. $q_{1}$, $q_{2}$ stands for $z$ and
$\alpha$, respectively, while $p_{1}$, $p_{2}$ for the associate momenta with
$z$, $\alpha$, respectively. $U^{J}$ is the liquid drop potential calculated
by two-center model\cite{sato} in addition with the rotational energy of the
system calculated with rigid body moment of inertia. $R_{i}$ is the random
force with Gaussian distribution,

%**********************figure 3 *************************

\begin{figure} [h]
\centerline{\psfig{file=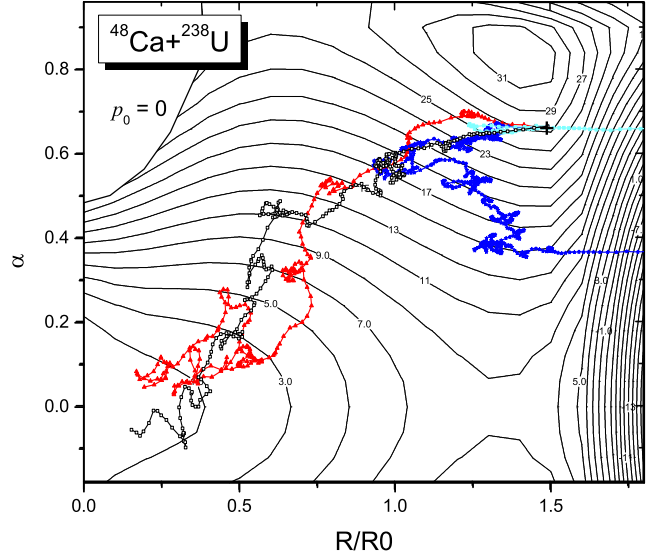,width=8.5cm}}
\vspace*{8pt}
\caption{Effect of the random force in the evolution from contact point to the
ground state. See text in more detail. }
\label{fig3}
\end{figure}

\begin{align*}
\left\langle R_{i}(t)\right\rangle  &  =0,\\
\left\langle R_{i}(t)R_{j}(t^{\prime})\right\rangle  &  =2\delta_{ij}
\delta(t-t^{\prime}).
\end{align*}
$g_{ij}$ is the strength of the random force which depends on the friction
tensor $\gamma_{ij}$ and temperature $T$ through the third line in Eq.(\ref{eq-5}), 
where $T^{J}=\sqrt{(E_{\rm c.m.} +Q-E_{\rm shell}-E_{\rm rot}^{J})/a}$ is defined in the case of 
compound ground state for each total angular momentum. 
The shell correction energy $E_{\rm{shell}}$ and
the nuclear mass are taken from P. M\"{o}ller's calculations\cite{moller}. The
level density parameter $a$ is taken approximately to be  $(A_{\rm p}+A_{\rm t})/10$ 
for massive nucleus. 

Examples of trajectories obtained are displayed in Fig. \ref{fig3}, with four 
trajectories starting from the same contact point with the same momentum and 
evolving in the same liquid drop energy surface. It shows that the random force 
plays very crucial role in the formation of compound nucleus. 
In Fig. \ref{fig3}, two samples form compound
nucleus and the others undergo a re-separation (quasi-fission). At the contact
point only initial momentum in $z$ degree of freedom is included because we
assume that $\alpha$ does not change in the approaching phase, and then no
initial momentum should be considered in the $\alpha$ degree of freedom.
(In reality, nucleon exchanges would occur, which gives rise to a distribution,
and maybe to initial momentum.)
Calculating $N$ trajectories for the same initial radial momentum $p_{0}$ and
counting the number of trajectories, which form compound nucleus, as
$N^{\prime}$, then a probability is given as
$$
F^{J}(p_{0},T^{J})=N^{\prime}/N.
$$
Because the radial momentum in the contact point is Gaussian distributed, 
centered at $\bar{p}_{0}^{J}$ due to a heating-up process by the 
dissipation-fluctuation dynamics,
$$
g^{J}(p_{0},\bar{p}_{0}^{J},T^{J})=\frac{1}{\sqrt{2\pi\mu T}}e^{-(p_{0}-\bar
{p}_{0}^{J})^{2}/(2\mu T)},
$$
then finally the formation probability takes the form,
\begin{equation}
P_{\rm{form}}^{J}(E_{\rm{c.m.}})=\int F^{J}(p_{0},T)g^{J}(p_{0},\bar
{p}_{0}^{J},T^{J})dp_{0}, \label{eq-6}
\end{equation}
which gives an example of the statistical connection between approaching phase 
and formation phase. Here $\bar{p}_{0}^{J}$, i.e., the average of $p_0$, is 
set to zero, according to the results by the surface friction model.  However 
it should be noticed that $\bar{p}_{0}^{J}=0$ does not always hold,
for example, in lighter systems, such as $^{100}$Mo + $^{100}$Mo. 
Inserting the results of Eq.(\ref{eq-3}) and Eq.(\ref{eq-6}) into
Eq.(\ref{eq-2}), the fusion probability for each partial wave and consequently
fusion cross section can be calculated.

\subsection{Fusion cross section}

%**********************figure 4*************************
\begin{figure} [b]
\centerline{\psfig{file=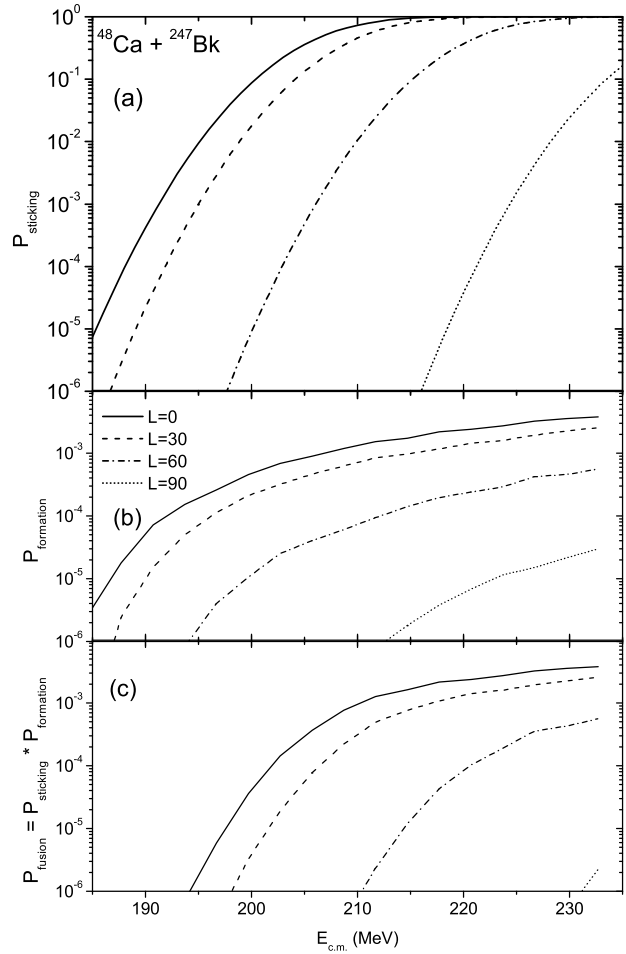,width=8.23cm}}
\vspace*{8pt}
\caption{The probabilities in the approaching phase, formation phase and their
product for $^{48}$Ca+$^{247}$Cm. (a) sticking probability, (b) formation
probability, (c) fusion probability (product of sticking and formation
probability).}
\label{fig-4}
\end{figure}

With the method described in Sec. 2, $P_{\rm{stick}}$ and $P_{\rm{form}}$
are calculated for different partial waves in reactions $^{48}$Ca +
$^{243-251}$Bk. As an example, the results of $P_{\rm{stick}}$,
$P_{\rm{form}}$ and $P_{\rm{fusion}}$ for $^{48}$Ca + $^{247}$Bk are shown in Fig.
\ref{fig-4}. It is obvious that the three probabilities increase with
increasing incident energy, while in the angular momentum direction, a
larger angular momentum gives smaller probabilities because the rotational
energies increase both the Coulomb barrier in approaching phase and liquid
drop saddle point in the formation phase.

\begin{figure} [h]
\centerline{\psfig{file=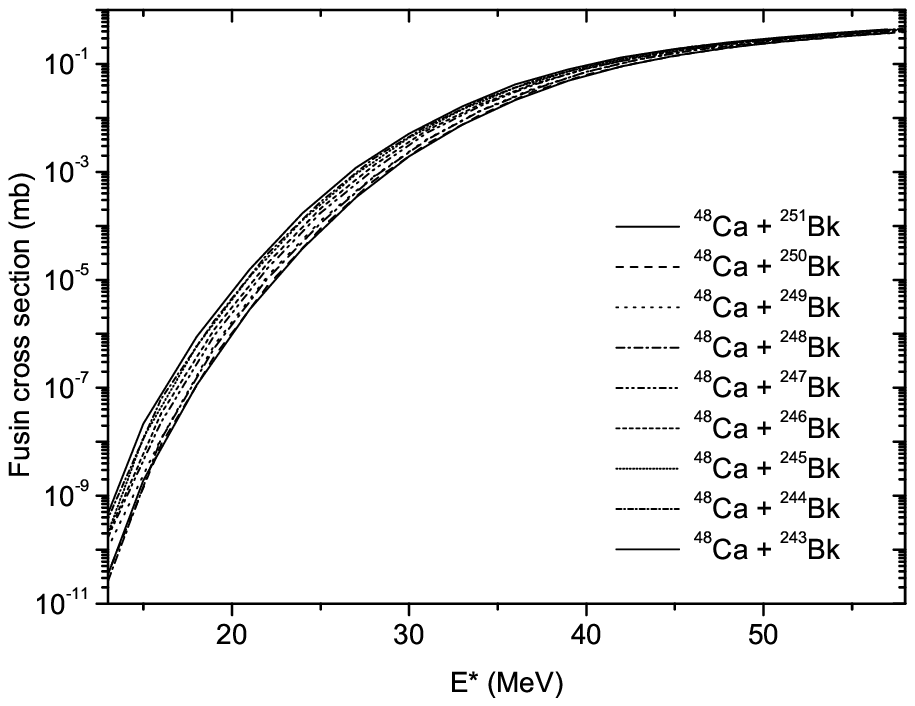,width=8.2cm}}
\vspace*{8pt}
\caption{Calculated fusion cross sections for $^{48}$Ca + $^{243-251}$Bk reactions.}
\label{fig-5}
\end{figure}

%**********************figure 6 *************************
\begin{figure} [h]
\centerline{\psfig{file=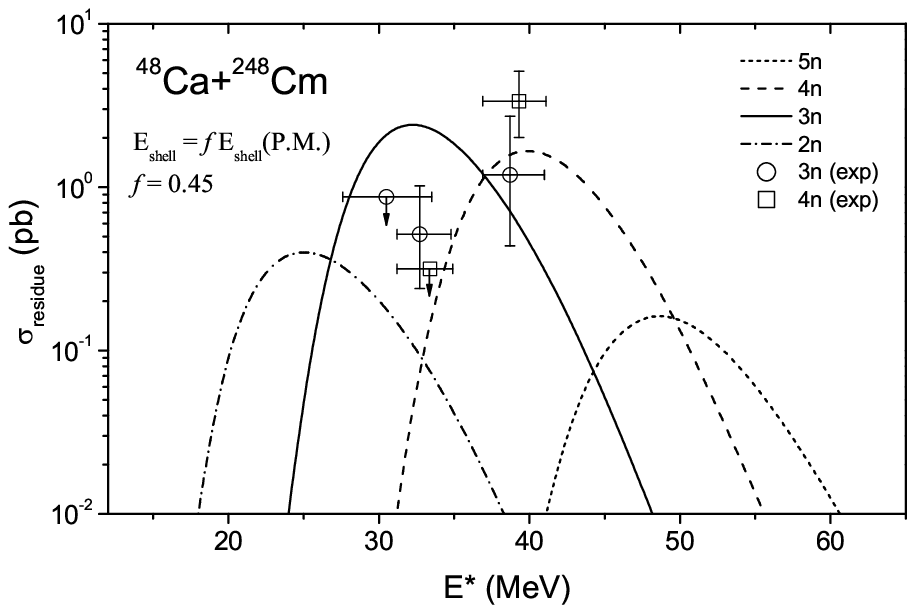,width=9.64cm}}
\vspace*{8pt}
\caption{Evaporation residue cross sections. The shell reduction parameter 
         $f$ is set to 0.45 to reproduce the data in Ref.[1]}
\label{fig-6}
\end{figure}

After systematic calculation of $P_{\rm stick}^J(E_{\rm c.m.})$ and 
$P_{\rm form}^J(E^*)$ for each reaction and taking Eq.(\ref{eq-22}) into 
account, the fusion cross sections for $^{48}$Ca + $^{243-251}$Bk are 
calculated and shown in Fig. \ref{fig-5}. Since the fusion cross section 
mainly depends on the bulk properties, such as Coulomb potential, liquid drop 
potential, the variation of $\sigma_{\rm{fusion}}$ for different targets is not 
very large. For example, at $E^{\ast}$ = 20 MeV, the ratio of $\sigma_{\rm{fusion}}$ 
between targets $^{243}$Bk and $^{251}$Bk is only 5.85.

%**********************figure 7 *************************
\begin{figure}[h]
\centerline{\psfig{file=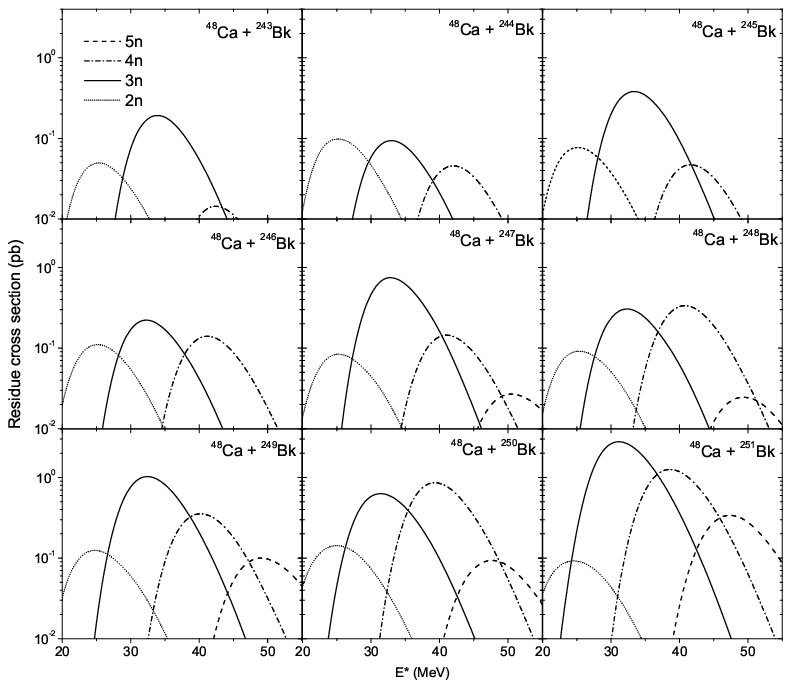,width=12.4cm}}
\vspace*{8pt}
\caption{Systematic calculation of residue cross sections for $^{48}$Ca +
$^{243-251}$Bk. The dotted line, solid line, dash-dotted line and dashed line
stands for $2n$, $3n$, $4n$, $5n$ evaporation channels, respectively.}
\label{fig-7}
\end{figure}

%**********************figure 8 *************************
\begin{figure}[h]
\centerline{\psfig{file=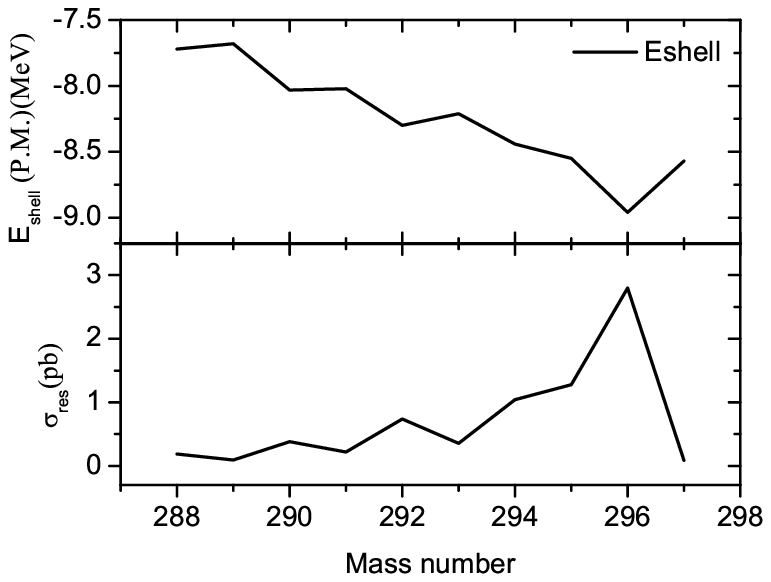,width=7.75cm}}
\vspace*{8pt}
\caption{The relation between the maximum residue cross sections and corresponding
shell correction energies taken from Ref.[20].}
\label{fig-8}
\end{figure}

\section{Calculation of residue cross sections and discussions}

In order to calculate the residue cross sections, the {\small HIVAP} code, based on
standard evaporation decay theory, is adopted. Before systematic calculations
for Ca + Bk, we try to obtain some parameters by fitting the measured $3n$ 
and $4n$ evaporation residue cross sections for $^{48}$Ca + $^{248}$Cm\cite{oganessian}.
Actually the shell correction energies are the most crucial quantities in
residue calculations, because they effectively give the fission barrier for
SHE. They are not yet firmly predicted, thus we may use the shell correction 
energy from Ref.\cite{moller} but with a free reduction parameter $f$, namely, 
$E_{\rm shell}$ = $f\cdot E_{\rm shell0}$. The factor $f$ is a reduction of the 
shell correction energy, and thus, that of the fission barrier in SHEs. Therefore, 
it gives rise to reductions of the absolute values of residue cross sections, but 
does not change general feature of the excitation functions, i.e., peak 
positions etc., though decreasing slopes in higher energies are a little affected. 
The introduction of the factor $f$, thus, is appropriate for predictions 
of residue cross sections. Using fusion probability for $^{48}$Ca + $^{248}$Cm
and the {\small HIVAP} code, $f$ is determined to be 0.45. 
Results are shown in Fig. \ref{fig-6}. It is seen that the residue
cross section of the $3n$ and $4n$ channels are well reproduced with the value of $f$ = 0.45.  
Using the same value of $f$, the calculated residue cross sections for $^{48}$Ca + $^{244}$Pu 
are also coincide with the experimental data. Since the targets of the reactions we study have 
only one more proton, the value of $f$ should work also in Ca+Bk case.

With all the ingredients that do not leave any ambiguity, systematic residue 
cross sections for $^{48}$Ca + $^{243-251}$Bk are calculated. The corresponding 
results are shown in Fig. \ref{fig-7}. It is easy to see that because the shell correction
energy gives rise to crucial effect on the fission barrier, the variation of the
maximum residue cross section among different reactions is larger than that
for the fusion cross section. For $^{243}$Bk the maximum residue 
cross section is $\sigma_{\rm res}$ = 0.19 pb
in $3n$ evaporation channel, while for $^{251}$Bk the corresponding value is 2.8
pb, one order larger. In order to show the importance of the shell effects to
the residue cross section, the relation between maximum $\sigma_{\rm res}$ and shell
correction energy is given in Fig. \ref{fig-8}.

In the present study, the fission barrier in decay process is $B_{\rm{f}
}=B_{\rm{LD}}-E_{\rm{shell}}$.\ Since the LD fission barrier $B_{\rm{LD}}$ is 
very small in the reactions considered\cite{dahlinger}, the fission barrier
mainly depends on the shell correction energy. Therefore in Fig.
\ref{fig-8}, when the shell correction is decreasing with increasing mass
number, the maximum residue cross section for SHE is increasing. It shows
that the shell correction is very important in the prediction of SHE, and
consequently these predictions are very sensitive to the reliability of 
this parameter.

%*********************** TABLE I *************************
\begin{table}[h]
\tbl{The residue cross sections of 3n and 4n evaporation channels for
     the $^{48}$Ca induced reactions to synthesize $Z=117$.}
{\begin{tabular}[c]{c|c|cc|cc}\hline\hline
Target & Lifetime & \multicolumn{2}{|c|}{3n} & \multicolumn{2}{|c}{4n}\\
& of target & $E_{\rm{c.m.}}$(MeV) & $\sigma_{\rm{res}}$(pb) & 
   $E_{\rm{c.m.}}$(MeV)  &  $\sigma_{\rm{res}}$(pb)\\\hline\hline
$^{243}$Bk & 4.50h & 207.5 & 0.19 & 215.9 & 0.015 \\ \hline
$^{244}$Bk & 4.35h & 206.1 & 0.09 & 215.1 & 0.045 \\ \hline
$^{245}$Bk & 4.94d & 206.0 & 0.38 & 214.4 & 0.047 \\ \hline
$^{246}$Bk & 1.80d & 204.5 & 0.22 & 213.4 & 0.14 \\ \hline
$^{247}$Bk & 1380y & 204.4 & 0.75 & 212.6 & 0.15 \\ \hline
$^{248}$Bk & 23.7h & 203.3 & 0.31 & 211.7 & 0.34 \\ \hline
$^{249}$Bk & 320d  & 203.3 & 1.04 & 211.0 & 0.35 \\ \hline
$^{250}$Bk & 3.2h  & 201.4 & 0.63 & 209.4 & 0.86 \\ \hline
$^{251}$Bk & 0.9h & 200.9 & 2.77 & 208.4 & 1.25 \\ \hline \hline
\end{tabular}}
\end{table}

In Table 1, the relatively larger residue cross sections of $^{48}$Ca-induced reactions 
to synthesize SHE are listed. The maximum residue cross section is for the reaction 
$^{48}$Ca + $^{251}$Bk where $\sigma_{\rm{res}}$ = 2.77 pb. However according to the
lifetime of the berkelium isotopes listed in Table 1, the lifetime of $^{251}$Bk
is too short to be a target since the experimental performance usually takes several
weeks to explore the SHE events. According to this factor, the optimum reactions to
synthesize $Z=117$ are $^{48}$Ca + $^{249}$Bk with $\sigma_{\rm res}$(3n) = 1.04 pb
at $E_{\rm c.m.}$ = 203.3 MeV and $^{48}$Ca + $^{247}$Bk with 
$\sigma_{\rm res}$(3n) = 0.75 pb at $E_{\rm c.m.}$ = 204.4 MeV. It should be mentioned 
that the absolute residue cross sections depend on the reduction factor $f$ of 
$E_{\rm shell}$ of the compound system and the reliability of 
$E_{\rm shell}$ itself. It has been confirmed that if the factor $f$ is changed, 
only the absolute values of $\sigma_{\rm res}$ are changed, while the relativity of the
$\sigma_{\rm res}$ between different reactions and the incident energies correspond to 
the maximum $\sigma_{\rm res}$ change with only negligible values. The result indicates
that the reaction $^{48}$Ca + $^{249}$Bk is optimum, no matter which value $f$ takes.

\section{Summary}

As a summary, the fusion reactions where $^{48}$Ca bombards berkelium isotopes
are studied with the two-step model, in which the fusion process is considered to
include two consecutive phases -- approaching phase and formation phase. In
the approaching phase, empirical formula by W. J. Swiatecki et al. is adopted
with the two parameters re-fitted to the superheavy systems. In the formation
phase, two dimensional Langevin equations are used to study the evolution of the
amalgamated system to the compound nucleus. Then, the {\small HIVAP} code 
is used to calculate the residue cross section. 
The results shows that an optimum reaction system is  $^{48}$Ca + $^{249}$Bk 
and that an optimum $E_{\rm{c.m.}}$ = 203.3 MeV for 3n residue cross section 1.04 pb.
Since the maximum value of the cross section is not extremely small and is within a 
capability of experiment nowadays, we expect experiment for the system be 
performed to result in synthesis of the new element with Z=117.
Of course, in principle, there are still other ways to synthesize $Z$ = 117, for
instance, Br + Pb, Se + Bi, Mn + U. However according to the two-step model, the residue
cross section would be smaller because of the larger fusion hindrance. 
To calculate residue cross sections, another code {\small KEWPIE 2} is now 
available, which is newly developed, carefully taking into account special
features in heavy and superheavy region\cite{marchix}.
In future, by using the new code {\small KEWPIE} and the two-step model, we will make a 
systematic prediction for heavier elements, not only with the hot fusion path, 
but also cold fusion path.  

\section*{Acknowledgements}

The present work has been supported by Natural Science Foundation of China and
Natural Science Foundation of Zhejiang Province under the grant Nos. 10675046
and Y605476, respectively, and by JSPS grant No. 18540268. The authors 
also acknowledge supports and hospitality by RCNP
Osaka University, GANIL, and Huzhou Teachers College, which enable us to 
continue the collaboration.


\begin{thebibliography}{99}

\bibitem {oganessian}Yu. Ts. Oganessian, V. K. Utyonkov, Yu. V. Lobanov et al., 
    Phys. Rev. C {\bf 74}, 044602 (2006); Phys. Rev. C {\bf 70}, 064609 (2004); 
    Phys. Rev. C {\bf 69}, 021601 (2004).

\bibitem {cwshen}C. W. Shen, G. Kosenko, Y. Abe, Phys. Rev. C {\bf 66}, 061602R (2002);
     B. Bouriquet et al., Eur. Phys. J. A {\bf 22}, 9 (2004);
     Y. Abe et al., Phys. Atom. Nucl. {\bf 69}, 1101 (2006).

\bibitem {dns}W. Li, N. Wang, J. F. Li, et al., Euro. Phys. Lett. {\bf 64},
     750(2003).

\bibitem {qmd}N. Wang, X. Z. Wu, Z. X. Li, et al., Phys. Rev. C {\bf 74}, 044604(2006).

\bibitem {swiatecki2005}W. J. Swiatecki, K. Siwek-Wilczynska, and J. Wilczynski, Phys.
Rev. C {\bf 71}, 014602 (2005).

\bibitem {jdbao}Z. H. Liu, J. D. Bao, Phys. Rev. C {\bf 74}, 057602(2006).

\bibitem {abe-eurojp}Y. Abe, Eur. Phys. J. A {\bf 13}, 143 (2002). 

\bibitem {abe2000} Y. Abe, D. Boilley, B. G. Giraud et al., Phys. Rev. E {\bf 61}, 1125 (2000).

\bibitem {reisdorf} W. Reisdorf (private communication).

\bibitem {frobrich}P. Frobrich et al., Nucl. Phys. A {\bf 406}, 557 (1983).

\bibitem {swiatecki1981}W. J. Swiatecki, Phys. Scr. {\bf 24}, 113 (1981).

\bibitem {abe2002}Y. Abe et al., Prog. Theor. Phys. Suppl. {\bf 146}, 104(2002);
                  Y. Abe et al., Acta Phys. Pol. B {\bf 34}, 2091 (2003).

\bibitem {boilley2003} D. Boilley et al., Eur. Phys. J. A {\bf 18}, 627 (2003).

\bibitem {wilczynska}K. Siwek-Wilczynska, E. Siemaszko and J. Wilczynski, 
         Acta Phys. Pol. B {\bf 33}, 451 (2002); K. Siwek-Wilczynska, J. Wilczynski,
         Phys. Rev. C {\bf 69}, 024611 (2004).

\bibitem {itkis}M. G. Itkis et al., Nuovo Cimento A {\bf 111}, 783 (1998).

\bibitem {boilley}D. Boilley et al., in preparation.

\bibitem {abe2008}Y Abe et al., Proc. Cluster08, to appear in Intern. J. Modern Phys. E (2008).

\bibitem {wada}T. Wada et al., Phys. Rev. Lett. {\bf 70}, 3538 (1993); 
       Y. Abe et al., Phys. Rep. {\bf 275}, 49 (1996).

\bibitem {sato}K. Sato et al., Z. Phys. A {\bf 290}, 145 (1979).

\bibitem {moller}P. M\"{o}ller et al., Atom. Data Nucl. Data Tables {\bf 59}, 185 (1995).

\bibitem {dahlinger}M. Dahlinger and D. Vermeulen, Nucl. Phys. A {\bf 376}, 94 (1982).

\bibitem {marchix}A. Marchix, Ph. D. thesis in Oct. 2007, Caen University, France.


\end{thebibliography}
\end{document}